\documentclass{PoS}

\title{An extension to the L\"uscher's finite volume method above inelastic threashold (formalism)}

\ShortTitle{A coupled channel extension to the L\"uscher's finite volume method}

\author{\speaker{Noriyoshi ISHII}\\
        Department of Physics, The University of Tokyo\\
        E-mail: \email{ishii@ribf.riken.jp}}
\author{for HAL-QCD Collaboration}
\author{\includegraphics[width=0.20\textwidth]{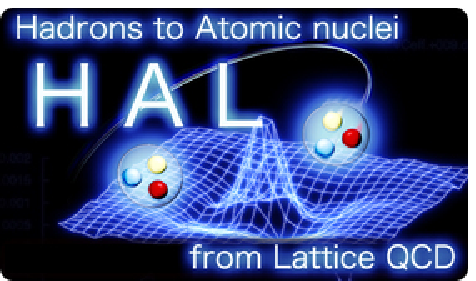}}

\abstract{An extension  of the L\"uscher's finite  volume method above
  inelastic thresholds is proposed.   It is fulfilled by extendind the
  procedure recently  proposed by  HAL-QCD Collaboration for  a single
  channel  system.   Focusing  on  the  asymptotic  behaviors  of  the
  Nambu-Bethe-Salpeter (NBS) wave  functions (equal-time) near spatial
  infinity,  a coupled  channel extension  of  effective Schr\"odinger
  equation  is   constructed  by  introducing   an  energy-independent
  interaction  kernel.  Because  the  NBS wave  functions contain  the
  information of  T-matrix at long distance, S-matrix  can be obtained
  by solving  the coupled channel effective  Schr\"odinger equation in
  the infinite volume.}

\FullConference{The XXVIII International Symposium on Lattice Field Theory, Lattice2010\\
		June 14-19, 2010\\
		Villasimius, Italy}

\newcommand{\Eq}[1]{Eq.~(\ref{#1})}
\newcommand{\Sect}[1]{Sect.~\ref{#1}}
\newcommand{\Ref}[1]{Ref.~\cite{#1}}

\begin{document}

\section{Introduction}
The standard  method to  calculate the scattering  phase shift  on the
lattice  is the  L\"uscher's finite  volume  method\cite{luscher}.  It
utilizes the  shift of  energy spectra in  the finite periodic  box to
calculate the  scattering phase.  The  formalism is restricted  to the
elastic scattering region.  One attempted to extend it above inelastic
thresholds \cite{liu}.   Although it can lead to  a constraint imposed
on  the  matrix elements  of  S-matrix,  S-matrix  elements cannot  be
obtained separately.

We  propose  a different  method  to  extend  it above  the  inelastic
threshold.   This is  fulfilled  by extending  the procedure  recently
proposed by  HAL-QCD Collaboration \cite{ishii, saoki}.
We   focus   on   the   asymptotic   behaviors   of   the   equal-time
Nambu-Bethe-Salpeter (NBS)  wave functions near  the spatial infinity,
where the information of T-matrix is contained.
We  construct a  coupled channel  effective Schr\"odinger  equation by
introducing an  energy-independent interaction kernel, so  that it can
generate all the NBS wave functions simultaneously.
Once such an  interaction kernel is constructed in  the finite volume,
S-matrix is obtained by  solving the effective Schrodinger equation in
the infinite  volume.  (The resulting  wave functions are  nothing but
the NBS wave  functions, which contain the information  of T-matrix in
their long distance part.)
In this paper, we use $N\Lambda$-$N\Sigma$ coupled system to formalize
the procedure.

Contents  are organized as  follows. In  \Sect{section.asymptotic}, we
consider  the  asymptotic  long  distance  behavior of  the  NBS  wave
functions for $N\Lambda$-$N\Sigma$ coupled system.
We will see that they contain T-matrix in their long distance part.
In   \Sect{section.effective_schrodinger},  we  extend   an  effective
Schr\"odinger equation  for the coupled channel  system by introducing
an energy-independent interaction kernel.
The extension  is performed so that  it can generate all  the NBS wave
functions simultaneously as solutions.
In \Sect{section.factorization}, we  derive the factorization formula,
which  plays a  key role  in constructing  an  effective Schr\"odinger
equation      in      a       coupled      channel      system      in
\Sect{section.effective_schrodinger}.

\section{Asymptotic behaviors of equal-time Nambu-Bethe-Salpeter (NBS)
  wave functions for a coupled channel system}
\label{section.asymptotic}
We consider $N\Lambda$-$N\Sigma$ coupled system with $m_{N}\simeq 940$
MeV, $m_{\Lambda}\simeq 1115$ MeV, and $m_{\Sigma}\simeq 1190$ MeV.
For notational simplicity, we treat them as bosons.
We are interested in the energy region
\begin{equation}
  m_N + m_{\Lambda}
  \le E \le
  m_{N} + m_{\Lambda} + m_{\pi},
  \label{eq.energy.region}
\end{equation}
which is  a combined region of  (i) elastic $N\Lambda$  region: $m_N +
m_{\Lambda} \le E  \le m_{N}+m_{\Sigma}$ and (ii) $N\Lambda$-$N\Sigma$
coupling region above the $N\Sigma$ threshold: $m_N + m_{\Sigma} \le E
\le m_{N} + m_{\Lambda} + m_{\pi}$.

We consider  two incoming states at  the same energy  $E$ as $|N(\vec
p)\Lambda(-\vec    p),in\rangle$     and    $|N(\vec    q)\Sigma(-\vec
q),in\rangle$. $\vec p$ and $\vec q$ denote the asymptotic momenta for
$N\Lambda$ and  $N\Sigma$ systems, respectively.  They  are related to
$E$ as
\begin{equation}
  E  = \sqrt{m_N^2 +  \vec p^2}  + \sqrt{m_{\Lambda}^2  + \vec  p^2} =
  \sqrt{m_N^2 + \vec q^2} + \sqrt{m_{\Sigma}^2 + \vec q^2}.
\end{equation}
We  define the  NBS wave  functions  for an  incoming state  $|N(\vec
p)\Lambda(-\vec p),in\rangle$ as
\begin{eqnarray}
  \psi_{N\Lambda; N\Lambda}(x_1,x_2; E)
  &\equiv&
  Z_{N}^{-1/2}
  Z_{\Lambda}^{-1/2}
  \langle 0 |
  T[N(x_1)\Lambda(x_2)]
  |N(\vec p)\Lambda(-\vec p),in\rangle
  \\
  \psi_{N\Sigma; N\Lambda}(x_1,x_2; E)
  &\equiv&
  Z_{N}^{-1/2}
  Z_{\Sigma}^{-1/2}
  \langle 0 |
  T[N(x_1)\Sigma(x_2)]
  |N(\vec p)\Lambda(-\vec p),in\rangle,
\end{eqnarray}
and  the   NBS  wave  functions   for  an  incoming   state  $|N(\vec
q)\Sigma(-\vec q),in\rangle$ as
\begin{eqnarray}
  \psi_{N\Lambda; N\Sigma}(x_1,x_2; E)
  &\equiv&
  Z_{N}^{-1/2}
  Z_{\Lambda}^{-1/2}
  \langle 0 |
  T[N(x_1)\Lambda(x_2)]
  |N(\vec q)\Sigma(-\vec q),in\rangle
  \\
  \psi_{N\Sigma; N\Sigma}(x_1,x_2; E)
  &\equiv&
  Z_{N}^{-1/2}
  Z_{\Sigma}^{-1/2}
  \langle 0 |
  T[N(x_1)\Sigma(x_2)]
  |N(\vec q)\Sigma(-\vec q),in\rangle,
\end{eqnarray}
where  $N(x)$,  $\Lambda(x)$ and  $\Sigma(x)$  denote local  composite
interpolating  fields  for nucleon,  $\Lambda$  and $\Sigma$  baryons,
respectively.
$Z_{N}$,  $Z_{\Lambda}$  and  $Z_{\Sigma}$  denote  the  normalization
factors   involved  in   the   limit  $N(x)\to   Z_N^{1/2}N_{out}(x)$,
$\Lambda(x)\to Z_{\Lambda}^{1/2}  \Lambda_{out}(x)$, and $\Sigma(x)\to
Z_{\Sigma}^{1/2} \Sigma_{out}(x)$, respectively, as $x_0 \to +\infty$.
By using the  reduction formula, these NBS wave  functions are related
to S-matrix as
\begin{eqnarray}
  \lefteqn{
    \langle N(p_1')\Lambda(p_2'),out|N(\vec p)\Lambda(-\vec p),in\rangle
  }
  \label{eq.S.NLNL}
  \\\nonumber
  &=&
  \mbox{disc.}
  +
  \int d^4 x_1 d^4 x_2
  e^{ip_1' x_1}\left(\square_1 + m_{N}^2\right)
  e^{ip_2' x_2}\left(\square_2 + m_{\Lambda}^2\right)
  \psi_{N\Lambda; N\Lambda}(x_1,x_2; E)
  \\
  \lefteqn{
    \langle N(q_1')\Sigma(q_2'),out|N(\vec p)\Lambda(-\vec p),in\rangle
  }
  \label{eq.S.NSNL}
  \\\nonumber
  &=&
  \hspace*{3em}
  \int d^4 x_1 d^4 x_2
  e^{iq_1' x_1}\left(\square_1 + m_{N}^2\right)
  e^{iq_2' x_2}\left(\square_2 + m_{\Sigma}^2\right)
  \psi_{N\Sigma; N\Lambda}(x_1,x_2; E),
\end{eqnarray}
and
\begin{eqnarray}
  \lefteqn{
    \langle N(p_1')\Lambda(p_2'),out|N(\vec q)\Sigma(-\vec q),in\rangle
  }
  \label{eq.S.NLNS}
  \\\nonumber
  &=&
  \hspace{3em}
  \int d^4 x_1 d^4 x_2
  e^{ip_1'x_1}\left(\square_1 + m_N^2\right)
  e^{ip_2'x_2}\left(\square_2 + m_{\Lambda}^2\right)
  \psi_{N\Lambda; N\Sigma}(x_1,x_2; E)
  \\
  \lefteqn{
    \langle N(q_1')\Sigma(q_2'),out| N(\vec q)\Sigma(-\vec q),in\rangle
  }
  \label{eq.S.NSNS}
  \\\nonumber
  &=&
  \mbox{disc.}
  +
  \int d^4 x_1 d^4 x_2
  e^{iq_1' x_1}\left(\square_1 + m_N^2 \right)
  e^{iq_2' x_2}\left(\square_2 + m_{\Sigma}^2\right)
  \psi_{N\Sigma; N\Sigma}(x_1, x_2; E),
\end{eqnarray}
where ``disc.''  stands for  disconnected terms, and $\square_i \equiv
\partial^2/\partial   t_i^2  -   \vec\nabla_i^2$   denotes  d'Alembert
operator.
These  relations lead  us  to the  following  asymptotic behaviors  of
equal-time restrictions  of NBS  wave functions near  spatial infinity
($|\vec x - \vec y| \to \infty$) as
\begin{eqnarray}
  \psi_{N\Lambda;N\Lambda}(\vec x - \vec y; E)
  &\equiv&
  \lim_{x_0\to +0}
  \psi_{N\Lambda;N\Lambda}(\vec x,x_0,\vec y,y_0=0; E)
  \label{eq.asymp.NLNL}
  \\\nonumber
  &\simeq&
  e^{i\vec p\cdot \vec r}
  +
  \frac{2}{E_N(p) + E_{\Lambda}(p)}
  \mathcal{T}_{N\Lambda;N\Lambda}(s)
  \frac{e^{ipr}}{pr}
  + \cdots
  \\
  \psi_{N\Sigma;N\Lambda}(\vec x - \vec y; E)
  &\equiv&
  \lim_{x_0\to +0}
  \psi_{N\Sigma;N\Lambda}(\vec x,x_0,\vec y,y_0=0; E)
  \label{eq.asymp.NSNL}
  \\\nonumber
  &\simeq&
  \hspace{2.5em}
  \frac{2}{E_{N}(q) + E_{\Sigma}(q)}
  \mathcal{T}_{N\Sigma ;N\Lambda}(s)
  \frac{e^{iqr}}{qr}
  + \cdots,
\end{eqnarray}
and
\begin{eqnarray}
  \psi_{N\Lambda;N\Sigma}(\vec x - \vec y; E)
  &\equiv&
  \lim_{x_0\to +0}
  \psi_{N\Lambda;N\Sigma}(\vec x,x_0,\vec y,y_0=0; E)
  \label{eq.asymp.NLNS}
  \\\nonumber
  &\simeq&
  \hspace{2.5em}
  \frac{2}{E_{N}(p) + E_{\Lambda}(p)}
  \mathcal{T}_{N\Lambda;N\Sigma}(s)
  \frac{e^{ipr}}{pr}
  + \cdots
  \\
  \psi_{N\Sigma;N\Sigma}(\vec x - \vec y; E)
  &\equiv&
  \lim_{x_0\to +0}
  \psi_{N\Sigma;N\Sigma}(\vec x,x_0,\vec y,y_0=0; E)
  \label{eq.asymp.NSNS}
  \\\nonumber
  &\simeq&
  e^{i\vec q\cdot \vec r}
  +
  \frac{2}{E_{N}(q) + E_{\Sigma}(q)}
  \mathcal{T}_{N\Sigma ;N\Sigma}(s)
  \frac{e^{iqr}}{qr}
  + \cdots,
\end{eqnarray}
where  $\mathcal{T}$  denotes the  on-shell  T-matrix, i.e.,  $\langle
f,out|i,in\rangle   =   \mathbb{I}    +   i(2\pi)^4   \delta^4(P_f   -
P_i)\mathcal{T}_{f,i}$,  $E_{N}(p) \equiv  \sqrt{m_N^2  + \vec  p^2}$,
$E_{\Lambda}(p)    \equiv    \sqrt{m_{\Lambda}^2    +   \vec    p^2}$,
$E_{\Sigma}(p)  \equiv \sqrt{m_{\Sigma}^2  + \vec  p^2}$, and  $\vec r
\equiv \vec x - \vec y$.

Derivations of  these asymptotic behaviors  of NBS wave  functions are
similar to single-channel case \cite{saoki,lin,cppacs}.
Here,  we give  a brief  example of  $\psi_{N\Lambda;N\Lambda}(\vec r;
\vec p)$ case.
\begin{eqnarray}
  \psi_{N\Lambda;N\Lambda}(\vec r; E)
  &=&
  Z_{N}^{-1/2}
  Z_{\Lambda}^{-1/2}
  \langle 0| N(\vec r) \Lambda(\vec 0) | N(\vec p)\Lambda(-\vec p),in\rangle
  \label{eq.NLNL}
  \\\nonumber
  &=&
  Z_{N}^{-1/2}
  Z_{\Lambda}^{-1/2}
  \int
  \frac{d^3 k}{(2\pi)^3\, 2 E_{N}(k)}
  \langle 0| N(\vec r) | N(\vec k)\rangle
  \langle N(\vec k)| \Lambda(0) | N(\vec p)\Lambda(-\vec p),in\rangle
  +
  I(\vec r),
\end{eqnarray}
where $I(\vec r)$  is referred to as the  inelastic contribution, which
corresponds  to  contributions  not  associated  with  single  nucleon
intermediate state.
Since $I(\vec  r)$ does not  propagate long distance, we  will neglect
this term below.
Because of \Eq{eq.S.NLNL}, we have an equality
\begin{eqnarray}
  \lefteqn{
    \langle N(k_1)|\Lambda(0)|N(p_1)\Lambda(p_2),in\rangle
  }
  \\\nonumber
  &=&
  \langle 0|
  \Lambda(0) a_{N,in}(k_1)
  |N(p_1)\Lambda(p_2),in\rangle
  \\\nonumber
  &&
  +iZ_{N}^{-1/2}
  \int d^4 x_1 e^{ik_1 x_1}
  \left(\square_1 + m_N^2\right)
  \langle 0| T[N(x_1)\Lambda(0)]|N(p_1)\Lambda(p_2),in\rangle
  \\\nonumber
  &=&
  Z_{\Lambda}^{1/2}
  \,
  2E_{N}(p_1)
  \,
  (2\pi)^3 \delta^3(k_1 - p_1)
  +Z_{\Lambda}^{1/2}
  \frac{\mathcal{T}(N(k_1)\Lambda(p_1 + p_2 - k_1);N(p_1)\Lambda(p_2))}
       {-(k_1-p_1-p_2)^2 + m_{\Lambda}^2 - i\epsilon},
\end{eqnarray}
which  is  used to  replace  the second  factor  in  the integrand  of
\Eq{eq.NLNL} as
\begin{eqnarray}
  \psi_{N\Lambda;N\Lambda}(\vec r; E)
  &=&
  e^{i\vec p\cdot \vec r}
  +
  \int \frac{d^3 k}{(2\pi)^3\,2E_N(k)}
  \\\nonumber
  &&\times
  \frac1{E_{\Lambda}(k)-E_{N}(k) + E_{N}(p)+E_{\Lambda}(p)}
  \cdot
  \frac{\mathcal{T}(
    N(\vec k)\Lambda(-\vec k);
    N(\vec p)\Lambda(-\vec p)
    )
  }{
    E_{\Lambda}(k) + E_{N}(k) - E_{\Lambda}(p) - E_{N}(p) - i\epsilon
    }
  e^{i\vec k\cdot\vec r}.
\end{eqnarray}
\Eq{eq.asymp.NLNL} is  arrived at by performing  the integration.  The
integration  can  be carried  out  by  noticing  that the  propagating
degrees   of  freedom   come   from  the   pole  contribution,   i.e.,
$E_{\Lambda}(k) + E_{N}(k) \simeq E_{\Lambda}(p) + E_{N}(p)$,
where  the off-shell  $\mathcal{T}$ can  be replaced  by  the on-shell
$\mathcal{T}$.
The   derivations   of   \Eq{eq.asymp.NSNL},  \Eq{eq.asymp.NLNS}   and
\Eq{eq.asymp.NSNS} are similar.

\section{An extension of effective Schr\"odinger equation
  for a coupled channel system}
\label{section.effective_schrodinger}
To  construct  an  effective  Schr\"odinger equation,  we  define  the
following $K$ functions by  multiplying Helmholtz operators on the NBS
wave functions as
\begin{eqnarray}
  K_{N\Lambda; N\Lambda}(\vec x; E)
  &\equiv&
  \left(\triangle + \vec p^2 \right)
  \psi_{N\Lambda; N\Lambda}(\vec x; E)
  \\\nonumber
  K_{N\Sigma; N\Lambda}(\vec x; E)
  &\equiv&
  \left(\triangle + \vec q^2 \right)
  \psi_{N\Sigma; N\Lambda}(\vec x; E),
\end{eqnarray}
and
\begin{eqnarray}
  K_{N\Lambda; N\Sigma}(\vec x; E)
  &\equiv&
  \left(\triangle + \vec p^2 \right)
  \psi_{N\Lambda; N\Sigma}(\vec x; E)
  \\\nonumber
  K_{N\Sigma; N\Sigma}(\vec x; E)
  &\equiv&
  \left(\triangle + \vec q^2 \right)
  \psi_{N\Sigma; N\Sigma}(\vec x; E).
\end{eqnarray}
These  Helmholtz  operators  filter  out the  propagating  degrees  of
freedom in the  asymptotic forms of NBS wave  functions.  As a result,
these  $K$ functions become  localized objects,  i.e., $K$  is nonzero
only within the range of interaction.
We factorize these $K$ functions as
\begin{eqnarray}
  \lefteqn{
    \left[
      \begin{array}{ll}
	K_{N\Lambda; N\Lambda}(\vec x; E) &
	K_{N\Lambda; N\Sigma }(\vec x; E) \\
	K_{N\Sigma;  N\Lambda}(\vec x; E) &
	K_{N\Sigma;  N\Sigma }(\vec x; E) \\
      \end{array}
      \right]
  }
  \label{eq.factorization}
  \\\nonumber
  &=&
  \int d^3 y
  \left[
    \begin{array}{ll}
      U_{N\Lambda; N\Lambda}(\vec x, \vec y) &
      U_{N\Lambda; N\Sigma }(\vec x, \vec y) \\
      U_{N\Sigma;  N\Lambda}(\vec x, \vec y) &
      U_{N\Sigma;  N\Sigma }(\vec x, \vec y) \\
    \end{array}
    \right]
  \cdot
  \left[
    \begin{array}{ll}
      \psi_{N\Lambda; N\Lambda}(\vec y; E) &
      \psi_{N\Lambda; N\Sigma }(\vec y; E) \\
      \psi_{N\Sigma;  N\Lambda}(\vec y; E) &
      \psi_{N\Sigma;  N\Sigma }(\vec y; E) \\
    \end{array}
    \right],
\end{eqnarray}
with an $E$-independent kernel $U$.
(The     proof     of     the     factorization    is     given     in
\Sect{section.factorization}.)
Since $K$ is localized, $U$ has to be localized as well.

If such a factorization is possible, we have
\begin{eqnarray}
  \lefteqn{
    \left[
      \begin{array}{ll}
	\left(\triangle + \vec p^2\right)\psi_{N\Lambda; N\Lambda}(\vec x; E) &
	\left(\triangle + \vec p^2\right)\psi_{N\Lambda; N\Sigma }(\vec x; E) \\
	\left(\triangle + \vec q^2\right)\psi_{N\Sigma;  N\Lambda}(\vec x; E) &
	\left(\triangle + \vec q^2\right)\psi_{N\Sigma;  N\Sigma }(\vec x; E) \\
      \end{array}
      \right]
  }
  \\\nonumber
  &=&
  \int d^3 y
  \left[
    \begin{array}{ll}
      U_{N\Lambda; N\Lambda}(\vec x, \vec y) &
      U_{N\Lambda; N\Sigma }(\vec x, \vec y) \\
      U_{N\Sigma;  N\Lambda}(\vec x, \vec y) &
      U_{N\Sigma;  N\Sigma }(\vec x, \vec y) \\
    \end{array}
    \right]
  \cdot
  \left[
    \begin{array}{ll}
      \psi_{N\Lambda; N\Lambda}(\vec y; E) &
      \psi_{N\Lambda; N\Sigma }(\vec y; E) \\
      \psi_{N\Sigma;  N\Lambda}(\vec y; E) &
      \psi_{N\Sigma;  N\Sigma }(\vec y; E) \\
    \end{array}
    \right].
\end{eqnarray}
This implies that  any linear combinations of NBS  wave functions with
arbitrary complex numbers $\alpha$ and $\beta$ as
\begin{equation}
  \left\{
  \begin{array}{lll}
  \psi_{N\Lambda}(\vec x; E)
  &\equiv
  &\alpha
  \psi_{N\Lambda; N\Lambda}(\vec x; E)
  +
  \beta
  \psi_{N\Lambda; N\Sigma}(\vec x; E)
  \\[0.5ex]\displaystyle
  \psi_{N\Sigma}(\vec x; E)
  &\equiv
  &\alpha
  \psi_{N\Sigma; N\Lambda}(\vec x; E)
  +
  \beta
  \psi_{N\Sigma; N\Sigma}(\vec x; E)
  \end{array}
  \right.
\end{equation}
satisfy
\begin{equation}
  \left[
    \begin{array}{l}
      \left(\triangle + \vec p^2\right)\psi_{N\Lambda}(\vec x; E) \\
      \left(\triangle + \vec q^2\right)\psi_{N\Sigma }(\vec x; E)
    \end{array}
    \right]
  =
  \int d^3 y
  \left[
    \begin{array}{ll}
      U_{N\Lambda; N\Lambda}(\vec x, \vec y) &
      U_{N\Lambda; N\Sigma }(\vec x, \vec y) \\
      U_{N\Sigma;  N\Lambda}(\vec x, \vec y) &
      U_{N\Sigma;  N\Sigma }(\vec x, \vec y) \\
    \end{array}
    \right]
  \cdot
  \left[
    \begin{array}{l}
      \psi_{N\Lambda}(\vec y; E) \\
      \psi_{N\Sigma }(\vec y; E) 
    \end{array}
    \right],
  \label{eq.effective.schrodinger}
\end{equation}
which  serves  as  an  coupled  channel  extension  of  the  effective
Schr\"odinger equation.

Several comments are in order.

(i)  \Eq{eq.effective.schrodinger}  does not  depend  on a  particular
choice  of  boundary condition,  since  $\alpha$  and  $\beta$ can  be
arbitrarily  chosen.  This  is a  desirable property,  which  makes it
possible to determine the interaction  kernel $U$ in the finite volume.
(The volume has to be sufficiently  large compared to the range of the
interaction.)
Once $U$ is determined in  the finite volume, S-matrix can be obtained
by  solving  \Eq{eq.effective.schrodinger}  in  the  infinite  volume.
(\Eq{eq.effective.schrodinger}   generates  NBS   wave   functions  as
solutions,  which contain the  information of  T-matrix in  their long
distance part.)

(ii) The interaction kernel $U$  is most generally a non-local object.
As is  described in \Ref{saoki}, a  practical way to  construct such a
non-local  interaction  kernel  $U$  is  to  rely  on  the  derivative
expansion, which makes it possible  to construct $U$ order by order in
the  derivative  expansion  by  using  a finite  number  of  NBS  wave
functions of varying energy $E$ and angular momentum $L$.
The  most  challenging  step  in  fulfilling  this  procedure  is  the
construction  of   NBS  wave   functions  at  excited   energies  with
variational method. This turns out to be feasible.  (See \Ref{sasaki},
where        this         procedure        is        applied        to
$\Lambda\Lambda$-$N\Xi$-$\Sigma\Sigma$ coupled system.)

(iii) In  the non-relativistic  limit, $\vec p^2$  and $\vec  q^2$ are
approximately  related  as   $\vec  q^2/(2\mu_{N\Sigma})  \simeq  \vec
p^2/(2\mu_{N\Lambda})    +    m_{\Lambda}    -   m_{\Sigma}$,    where
$\mu_{N\Sigma}$  and  $m_{N\Lambda}$  denote  the reduced  masses  for
$N\Sigma$  and  $N\Lambda$  systems,  respectively.   In  this  limit,
\Eq{eq.effective.schrodinger} reduces to an eigenvalue problem used in
\Ref{liu}:
\begin{equation}
  \left[
    \begin{array}{cc}
      -\frac{\triangle}{2\mu_{N\Lambda}} + {\widetilde U}_{N\Lambda;N\Lambda} &
                                           {\widetilde U}_{N\Lambda;N\Sigma } \\
                                           {\widetilde U}_{N\Sigma ;N\Lambda} &
     m_{\Sigma} - m_{\Lambda}                                      
      -\frac{\triangle}{2\mu_{N\Sigma }} + {\widetilde U}_{N\Sigma ;N\Sigma } \\
    \end{array}
    \right]
  \cdot
  \left[
    \begin{array}{l}
      \psi_{N\Lambda}(\vec x; E) \\
      \psi_{N\Sigma }(\vec x; E) \\
    \end{array}
    \right]
  =
  E_{\rm NR}
  \left[
    \begin{array}{l}
      \psi_{N\Lambda}(\vec x; E) \\
      \psi_{N\Sigma }(\vec x; E) \\
    \end{array}
    \right],
\end{equation}
where
$\widetilde{U}_{N\Lambda;N\Lambda} \equiv U_{N\Lambda;N\Lambda}/(2\mu_{N\Lambda})$,
$\widetilde{U}_{N\Lambda;N\Sigma   }$   $\equiv$  $U_{N\Lambda;N\Sigma
}/(2\mu_{N\Lambda})$,
$\widetilde{U}_{N\Sigma      ;N\Lambda}$      $\equiv$     $U_{N\Sigma
  ;N\Lambda}/(2\mu_{N\Sigma })$,
$\widetilde{U}_{N\Sigma ;N\Sigma }$ $\equiv$ $U_{N\Sigma ;N\Sigma }/(2\mu_{N\Sigma })$, and
$E_{\rm NR}\equiv \vec p^2/(2\mu_{N\Lambda})$.

(iv)  In constructing  \Eq{eq.effective.schrodinger},  we assumed  the
orthogonality of NBS wave functions. If the violation of orthogonality
becomes severe, NBS wave functions should be orthogonalized before all
the above procedures begin.  The orthogonalization should be performed
without affecting  the long  distance behaviors, where  information of
T-matrix is contained.

\section{Proof of the factorization of K function}
\label{section.factorization}
We  assume  that  the   following  NBS  wave  functions  are  linearly
independent
\begin{equation}
  \left\{
  \left[
    \begin{array}{l}
      \psi_{N\Lambda; N\Lambda}(\vec x; E) \\
      \psi_{N\Sigma;  N\Lambda}(\vec x; E)
    \end{array}
    \right],
  \left[
    \begin{array}{l}
      \psi_{N\Lambda; N\Sigma }(\vec x; E) \\
      \psi_{N\Sigma;  N\Sigma }(\vec x; E)
    \end{array}
    \right]
  \right\}_{m_{N}+m_{\Lambda} \le E \le m_{N}+m_{\Lambda}+m_{\pi}}.
\end{equation}
There exists a dual basis
\begin{equation}
  \left\{
  \left[
    \widetilde{\psi}_{N\Lambda; N\Lambda}(\vec x; E),
    \widetilde{\psi}_{N\Lambda; N\Sigma} (\vec x; E)
    \right],
  \hspace{0.5em}\rule{0ex}{2.5ex}
  \left[
    \widetilde{\psi}_{N\Sigma; N\Lambda}(\vec x; E),
    \widetilde{\psi}_{N\Sigma; N\Sigma} (\vec x; E)
    \right]
  \right\}_{m_{N}+m_{\Lambda} \le E \le m_{N}+m_{\Lambda}+m_{\pi}},
\end{equation}
which serves as  a ``left inverse'' as an  integration operator in the
following sense:
\begin{equation}
  \int d^3 y
  \left[
    \begin{array}{ll}
      \widetilde{\psi}_{N\Lambda;N\Lambda}(\vec y; E') &
      \widetilde{\psi}_{N\Lambda;N\Sigma} (\vec y; E') \\
      \widetilde{\psi}_{N\Sigma; N\Lambda}(\vec y; E') &
      \widetilde{\psi}_{N\Sigma; N\Sigma} (\vec y; E') \\
    \end{array}
    \right]
  \cdot
  \left[
    \begin{array}{ll}
      \psi_{N\Lambda; N\Lambda}(\vec y; E) &
      \psi_{N\Lambda; N\Sigma} (\vec y; E) \\
      \psi_{N\Sigma;  N\Lambda}(\vec y; E) &
      \psi_{N\Sigma;  N\Sigma} (\vec y; E) \\
    \end{array}
    \right]
  =
  (2\pi) \delta(E - E').
\end{equation}
Now, the factorization can be verified in the following way:
\begin{eqnarray}
  \lefteqn{
    \left[
      \begin{array}{ll}
        K_{N\Lambda; N\Lambda}(\vec x; E) &
        K_{N\Lambda; N\Sigma }(\vec x; E) \\
        K_{N\Sigma;  N\Lambda}(\vec x; E) &
        K_{N\Sigma;  N\Sigma }(\vec x; E) \\
      \end{array}
      \right]
  }
  \\\nonumber
  &=&
  \int \frac{dE'}{2\pi}
  \left[
    \begin{array}{ll}
      K_{N\Lambda; N\Lambda}(\vec x; E') &
      K_{N\Lambda; N\Sigma }(\vec x; E') \\
      K_{N\Sigma;  N\Lambda}(\vec x; E') &
      K_{N\Sigma;  N\Sigma }(\vec x; E') \\
    \end{array}
    \right]
  \\\nonumber
  &&
  \hspace{3em}
  \times
  \int d^3 y
  \left[
    \begin{array}{ll}
      \widetilde{\psi}_{N\Lambda;N\Lambda}(\vec y; E') &
      \widetilde{\psi}_{N\Lambda;N\Sigma} (\vec y; E') \\
      \widetilde{\psi}_{N\Sigma; N\Lambda}(\vec y; E') &
      \widetilde{\psi}_{N\Sigma; N\Sigma} (\vec y; E') \\
    \end{array}
    \right]
  \cdot
  \left[
    \begin{array}{ll}
      \psi_{N\Lambda; N\Lambda}(\vec y; E) &
      \psi_{N\Lambda; N\Sigma} (\vec y; E) \\
      \psi_{N\Sigma;  N\Lambda}(\vec y; E) &
      \psi_{N\Sigma;  N\Sigma} (\vec y; E) \\
    \end{array}
    \right]
  \\\nonumber
  &=&
  \int d^3 y
  \left\{
  \int \frac{dE'}{2\pi}
  \left[
    \begin{array}{ll}
      K_{N\Lambda; N\Lambda}(\vec x; E') &
      K_{N\Lambda; N\Sigma }(\vec x; E') \\
      K_{N\Sigma;  N\Lambda}(\vec x; E') &
      K_{N\Sigma;  N\Sigma }(\vec x; E') \\
    \end{array}
    \right]
  \cdot
  \left[
    \begin{array}{ll}
      \widetilde{\psi}_{N\Lambda;N\Lambda}(\vec y; E') &
      \widetilde{\psi}_{N\Lambda;N\Sigma} (\vec y; E') \\
      \widetilde{\psi}_{N\Sigma; N\Lambda}(\vec y; E') &
      \widetilde{\psi}_{N\Sigma; N\Sigma} (\vec y; E') \\
    \end{array}
    \right]
  \right\}
  \\\nonumber
  &&
  \hspace{3em}
  \times
  \left[
    \begin{array}{ll}
      \psi_{N\Lambda; N\Lambda}(\vec y; E) &
      \psi_{N\Lambda; N\Sigma} (\vec y; E) \\
      \psi_{N\Sigma;  N\Lambda}(\vec y; E) &
      \psi_{N\Sigma;  N\Sigma} (\vec y; E) \\
    \end{array}
    \right].
\end{eqnarray}
We  arrive  at  the  factorization  formula  \Eq{eq.factorization}  by
defining an interaction kernel $U$ as
\begin{eqnarray}
  \lefteqn{
    \left[
      \begin{array}{ll}
        U_{N\Lambda; N\Lambda}(\vec x,\vec y) &
        U_{N\Lambda; N\Sigma }(\vec x,\vec y) \\
        U_{N\Sigma;  N\Lambda}(\vec x,\vec y) &
        U_{N\Sigma;  N\Sigma }(\vec x,\vec y) \\
      \end{array}
      \right]
  }
  \\\nonumber
  &\equiv&
  \int \frac{dE'}{2\pi}
  \left[
    \begin{array}{ll}
      K_{N\Lambda; N\Lambda}(\vec x; E') &
      K_{N\Lambda; N\Sigma }(\vec x; E') \\
      K_{N\Sigma;  N\Lambda}(\vec x; E') &
      K_{N\Sigma;  N\Sigma }(\vec x; E') \\
    \end{array}
    \right]
  \cdot
  \left[
    \begin{array}{ll}
      \widetilde{\psi}_{N\Lambda;N\Lambda}(\vec y; E') &
      \widetilde{\psi}_{N\Lambda;N\Sigma} (\vec y; E') \\
      \widetilde{\psi}_{N\Sigma; N\Lambda}(\vec y; E') &
      \widetilde{\psi}_{N\Sigma; N\Sigma} (\vec y; E') \\
    \end{array}
    \right].
\end{eqnarray}
Note that, due to the  integration of $E'$, the interaction kernel $U$
does not depend on the energy $E$.

\section{Summary}
\label{section.summary}
We have proposed an extension  of the L\"uscher's finite volume method
above the inelastic threshold.
We have considered $N\Lambda$-$N\Sigma$ coupled system as an example.
We have seen that equal-time Nambu-Bethe-Salpeter (NBS) wave functions
contain the  information of  T-matrix in their long distance part.
We have constructed an  effective Schr\"odinger equation for a coupled
channel  system  by   introducing  an  energy-independent  interaction
kernel, so  that it  can generate  all the NBS  wave functions  in the
energy region $m_{N}+m_{\Lambda} \le E \le m_{N}+m_{\Lambda}+m_{\pi}$.
Since this interaction  kernel is localized, it can  be constructed by
using lattice QCD calculation in a finite volume.
Once it is constructed in  the finite volume, S-matrix can be obtained
by  solving  the  effective  Schr\"odinger equation  in  the  infinite
volume.

It is interesting  to apply this method to  the hyperon systems, where
only a  limited amount of  experimental information is  available. The
first       attempt      to       apply      this       method      to
$\Lambda\Lambda$-$N\Xi$-$\Sigma\Sigma$  coupled system is  reported in
\Ref{sasaki},  where  the interaction  kernel  is  constructed at  the
leading order of the derivative  expansion by using NBS wave functions
obtained by the variational method.

\section*{Acknowledgments}
This  work was  supported  in  part by  Grant-in-Aid  of the  Japanese
Ministry of  Education, Science, Sport and Culture  (No. 22540268) and
in part by a Grand-in-Aid for Specially Promoted Research (13002001).

\end{document}